\documentclass[aps,prd,twocolumn,nofootinbib,floatfix,showpacs]{revtex4} 

\usepackage{graphicx}

\usepackage{latexsym}
\usepackage[cp866]{inputenc}
\usepackage[english]{babel}   

\input epsf

\begin{document}

\title{\boldmath Study of the $f_2(1270)$ and $a_2(1320)$ resonances
in $\gamma^*(Q^2)\gamma$ collisions}
\author{N.~N.~Achasov, A.~V.~Kiselev, and G.~N.~Shestakov}
\affiliation{Laboratory of Theoretical Physics, S.~L.~Sobolev
Institute for Mathematics, 630090 Novosibirsk, Russia}


\begin{abstract}
We discuss studies of the $Q^2$ dependence of the $f_2(1270)$ and
$a_2 (1320)$ production cross sections in $\gamma^*(Q^2)\gamma $
collisions at current and coming into operation colliders with a
high luminosity. Changing the dominant helicity amplitude occurs in
the reactions $\gamma^*(Q ^2)\gamma\to f_2(1270)$ and $\gamma^*(Q^2)
\gamma\to a_2(1320)$ with increasing $Q^2$. This is caused by the
coming of the QCD asymptotics. It is shown that the transition to
the asymptotic behavior of QCD in the amplitudes $\gamma^*(Q^2)
\gamma\to f_2(1270),a_2(1320)$ is provided by the compensation of
the contributions of ground vector states $\rho$ and $\omega$ in
$Q^2$-channel with the contributions of their radial excitations.
\end{abstract}


\maketitle

Physics of two-photons collisions entering into the era of
ultra-high statistics gives unique opportunities to study the
internal (quark-gluon) structure of hadrons \cite{PDG14,ABB09,
ABB10,BGM14}. For example, the recent experiments of the Belle
Collaboration on the reactions $\gamma\gamma$\,$\to$\,$\pi^+\pi^-$
\cite{Mo07a,Mo07b}, $\gamma \gamma$\,$\to$\,$\pi^0\pi^0$
\cite{Ue08}, and $\gamma\gamma$\,$\to $\,$\pi^0\eta$ \cite{Ue09}
established conclusively the smallness of the two-photon widths of
the $f_0(980)$ and $a_0(980)$ resonances, which testifies in favor
of their four-quark structure
\cite{ADS82a,ADS82b,AS11}.\,\footnote{In 1982, the prediction
$\Gamma(f_0(980)\to\gamma \gamma)\approx \Gamma(a_0(980)\to\gamma
\gamma)\approx0.27$ keV  was done in the four-quark MIT bag model
\cite{ADS82a,ADS82b}. In 2014, the Particle Data Group cited in the
Review of Particle Physics the following data \cite{PDG14}:
$\Gamma(f_0(980)\to\gamma\gamma) \approx0.29$ keV and $\Gamma(a_0
(980)\to\gamma\gamma)\approx0.3$ keV, which is an order of magnitude
smaller than the $\gamma\gamma$ width of the tensor $q\bar q$ meson
$\Gamma (f_2(1270)\to\gamma\gamma)\approx3$ keV. The prediction of
the $q\bar q$ model $\Gamma(f_0(980)\to\gamma \gamma)/\Gamma(a_0
(980)\to\gamma\gamma)=25/9$ is excluded experimentally.}

The measurements of the two-photon widths of the light pseudoscalar
mesons $P=\pi^0$, $\eta$, $\eta'$ in $\gamma\gamma$ collisions
\cite{PDG14} and the transition form factors $F_{\gamma^*\gamma\to
P}(Q^2)$ in $\gamma^*(Q^2)\gamma$
collisions\,\footnote{$\gamma^*(Q^2)$ ($\gamma^*$ below) denotes the
photon with virtuality $-Q^2$.} performed by CELLO \cite{Beh91},
CLEO \cite{Gro98}, BaBar \cite{Aub09,San11} and Belle \cite{Ueh12}
Collaborations allowed to realize a critical test of QCD
calculations of the processes at large $Q^2$.

\begin{figure}  
\includegraphics[width=8.5cm]{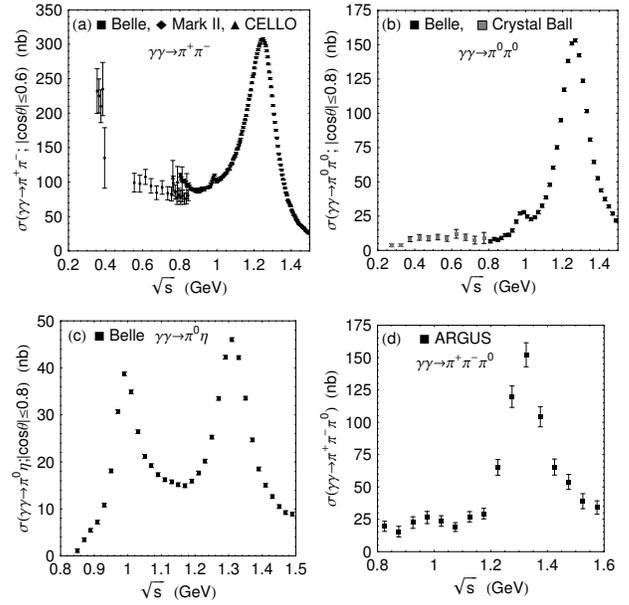}
\caption{\label{ggpipi} Cross sections of the reactions (a) $\gamma
\gamma \to\pi^+\pi^-$ \cite{Mo07a,Mo07b, BBG90,BCF92}, (b) $\gamma
\gamma \to\pi^0\pi^0$ \cite{Ue08,MAB90}, (c) $\gamma\gamma\to\pi^0
\eta$ \cite{Ue09}, and (d) $\gamma \gamma\to\pi^+\pi^-\pi^0$
\cite{ARG97} as functions of the invariant mass, $\sqrt{s}$, of the
final meson system. In plots (a)--(c) $\theta$ denotes the polar
angle of one of the outgoing mesons with respect to the incident
photon direction in the $\gamma\gamma$ center-of-mass system. The
reactions $\gamma\gamma\to\pi^0\pi^0$, plot (b), and
$\gamma\gamma\to\pi^+\pi^- \pi^0$, plot (d), seem more preferable in
the sense of the smallness of the physical background under the
$f_2(1270)$ and $a_2(1320)$ peaks, respectively.} \end{figure}

Production of classical tensor $q\bar q$ resonances by two real
photons proceeds very intensively: $f_2(1270)$ in the reactions
$\gamma\gamma$\,$\to$\,$\pi^+\pi^-$ \cite{Mo07a,Mo07b,BBG90,BCF92}
and $\gamma\gamma$\,$\to$\,$\pi^0\pi^0$ \cite{Ue08,MAB90} and
$a_2(1320)$ in the reactions $\gamma \gamma$\,$\to$\,$\pi^0\eta$
\cite{Ue09,AAB86} and $\gamma\gamma$\,$\to$\,$\pi^+\pi^-\pi^0$
\cite{BDG84,BBB90,ARG97, AAA97,AAA06} (see Fig. 1). This fact is a
good reason to start detailed investigations of the $Q^2$ dependence
of the $f_2(1270)$ and $a_2(1320)$ production cross sections in
$\gamma^*\gamma$ collisions at $e^+e^-$ colliders with a high
luminosity.\,\footnote{Currently, the maximum luminosity
$\approx2\cdot10^{34}$ cm$^{-2}\cdot c^{-1}$ is reached at the KEKB
$e^+e^-$ collider \cite{PDG14,BGM14}. The luminosity of
$8\cdot10^{35}$ cm$^{-2}\cdot c^{-1}$ is planned to have at the
SuperKEKB factory \cite{PDG14, ABB10}.}

We now turn to the detailed discussion.

In $\gamma\gamma$ collisions, the $f_2(1270)$ and $a_2(1320)$
resonances can be produced in the states with helicity
$\lambda$\,=\,0 and $\pm2$. Helicity $\lambda$ is defined in the
resonance rest frame, in which $\lambda$\,=\,$\lambda_1-\lambda_2$,
where $\lambda_1$ and $\lambda_2$ are the helicities of incoming
photons. According to the high-statistics measurements
\cite{Mo07b,Ue08,Ue09,BBG90,BCF92, MAB90,ARG97,AAA97,AAA06} the
fraction of the $f_2(1270)$ and $a_2(1320)$ production in states
with $\lambda$\,=\,$\pm2$ in $\gamma\gamma$ collisions is more than
95\%. 

This remarkable experimental fact of $\lambda$\,=\,$\pm2$ dominance
is naturally reproduced by the effective gauge-invariant Lagrangian,
describing the tensor meson production by two photons with opposite
helicities only \cite{ADS82b, AK86,AGKR13},
\begin{equation}\label{L-TGG}
L=g_{T\gamma\gamma}T_{\mu\nu}F_{\mu\sigma}F_{\nu\sigma}\,,
\end{equation}
where $F_{\mu\nu}=\partial_\mu A_\nu-\partial_\nu A_\mu$ is the
tensor of the electromagnetic field $A_\mu$, $T_{\mu\nu}$ is the
field of the tensor meson $T$\, ($T$\,=\,$f_2 (1270)$, $a_2(1320)$);
$T_{\mu\nu}=T_{\nu \mu} $, $T_{\mu\mu}=0$, $\partial_\mu
T_{\mu\nu}=0$; $g_{T\gamma \gamma}$ is the coupling constants of the
$T$ meson to the energy-momentum tensor of the electromagnetic
field.

Using Lagrangian (\ref{L-TGG}) one can unambiguously predict the
hierarchy of the $Q^2$ dependencies of the helicity amplitudes
$V^{(\lambda)}_{\lambda_1,\lambda_2}(T;\,s,Q^2)$\,=\,$V^{(-\lambda)
}_{-\lambda_1,-\lambda_2}(T;\,s,Q^2)$ describing the
$\gamma^*\gamma$\,$\to$\,$T$ vertices \cite{AK86,AKS15}:
\begin{equation}\label{FF2}
V^{(2)}_{1,-1}(T;\,s,Q^2)=V_T(s,Q^2)\left(1+\frac{Q^2}{s}\right),\ \
\end{equation}
\begin{equation}\label{FF1}
V^{(1)}_{1,0}(T;\,s,Q^2)=V_T(s,Q^2)\sqrt{\frac{Q^2}{2s}}\,\left(1+\frac{Q^2}{s}\right),
\end{equation}
\begin{equation}\label{FF0}
V^{(0)}_{1,1}(T;\,s,Q^2)=-V_T(s,Q^2)\frac{Q^2}{\sqrt{6}s}\left(1+\frac{Q^2}{s}\right).
\end{equation} Here $s=(q_1+q_2)^2$; $q_1$ and $q_2$ are the
four-momenta of the incident photons, $q^2_1=0$, $q^2_2=-Q^2$;
\begin{equation}\label{Norm1}
V_T(s,Q^2)=g_{T\gamma\gamma}\,s\,F_T(Q^2)/2,\quad F_T(0)=1,\ \
\end{equation}
\begin{equation}\label{Norm2}
g_{T\gamma\gamma}\,s=2V^{(2)}_{1,-1}(T;\,s,0)=\sqrt{320\,\pi\,
\sqrt{s}\,\Gamma_{T\to\gamma\gamma}(s)},
\end{equation}
and $F_T(Q^2)$ is the transition form factor which is common for all
vertices.

The vertex $V^{(1)}_{1,0}(T;\,s,Q^2)$ vanishes for $Q^2$\,$\to$\,0
as $\sqrt{Q^2}$. This is a consequence of gauge invariance. The
vertex $V^{(0)}_{1,1}(T;\,s,Q^2 )$ is proportional to $Q^2$ for
$Q^2$\,$\to$\,0 owing to a specific selection of the
$\gamma^*\gamma\, T$ interaction which consists with the
experimental fact of $\lambda$\,=\,$\pm2$ dominance in
$\gamma\gamma\to T$ transitions (see $V^{(2)}_{1,-1}(T; \,s,0)$ in
Eq. (\ref{Norm2})).

For small $Q^2$, the dominance of $V^{(2)}_{1,-1}(T;\,s,Q^2)$ over
$V^{(1)}_{1,0}(T;\,s,Q^2)$ and $V^{(0)}_{1,1}(T;\,s,Q^2)$ is
certainly maintained. However, for large $Q^2$ the situation changes
radically. Asymptotically
\begin{equation}\label{AFF2}
V^{(2)}_{1,-1}(T;\,s,Q^2)\sim F_T(Q^2)\,Q^2,
\end{equation}
\begin{equation}\label{AFF1}
V^{(1)}_{1,0}(T;\,s,Q^2)\sim F_T(Q^2)\,Q^3,
\end{equation}
\begin{equation}\label{AFF0}
V^{(0)}_{1,1}(T;\,s,Q^2)\sim F_T(Q^2)\,Q^4
\end{equation}
and the $\gamma^*\gamma\to T$ vertex with $\lambda$\,=\,0 becomes
dominant.

From the parton model considerations \cite{KWZ74} and the QCD
analysis of hard exclusive processes \cite{CZh84,BG85} it follows
that for large $Q^2$ the tensor meson production amplitude with zero
helicity (in the $\gamma^*\gamma$ center-of-mass system) should tend
to the constant value (with logarithmic accuracy), and other
amplitudes should be suppressed by powers of $Q^2$. This implies
that $F_T(Q^2)\sim1/Q^4$ for large $Q^2$. In the generalized vector
meson dominance model (GVDM) such an asymptotic behavior is provided
by the compensation in $Q^2$-channel of the contributions of ground
and excited states of vector mesons $V=\rho,\omega,\phi$,\,
$V'=\rho',\omega',\phi'$,\, $V''=\rho'',\omega'',\phi''$, etc.
\cite{AKS15}.

It is interesting to find out, at least roughly, how fast the
angular distributions can vary with $Q^2$ in the reactions
$\gamma^*\gamma\to f_2(1270)\to\pi\pi$,\, $\gamma^*\gamma\to
a_2(1320)\to\pi^0\eta$,\, and\, $\gamma^*\gamma\to
a_2(1320)\to\rho^\pm\pi^\mp\to\pi^+\pi^- \pi^0$ for $0<Q^2<40$
GeV$^2$ (in the case of the processes $\gamma^*\gamma\to\pi^0$,
$\eta$, $\eta'$ the asymptotic regime apparently occurs near 40
GeV$^2$).

We put $m_\rho$\,=\,$ m_\omega$, $m_{\rho'}$\,=\,$m_{\omega'}$,
etc., and will consider that the resonance $f_2(1270)$ does not
contain strange valent quarks (as $\omega$, $\omega'$, etc.). Then,
in GVDM, a simplest expression for $F_T(Q^2)$ with the required
asymptotic behavior has the form
\begin{equation}\label{FFQ2}
F_T(Q^2)=\frac{1}{(1+Q^2/m^2_\rho)(1+Q^2/m^2_{\rho'})}.
\end{equation} Figure 2 shows the $Q^2$ dependencies of the normalized
vertex functions $2|V^{(\lambda)}_{\lambda_1,
\lambda_2}(T;\,m^2_T,Q^2)|/(g_{T\gamma\gamma}m^2_T)$ calculated
according Eqs. (\ref{FF2})--(\ref{Norm2}) and (\ref{FFQ2}) at
$m_\rho=0.775$ GeV, $m_{\rho'}=1.465$ GeV, and $s=m^2_T$.
\begin{figure}  \begin{center}
\includegraphics[width=6.5cm]{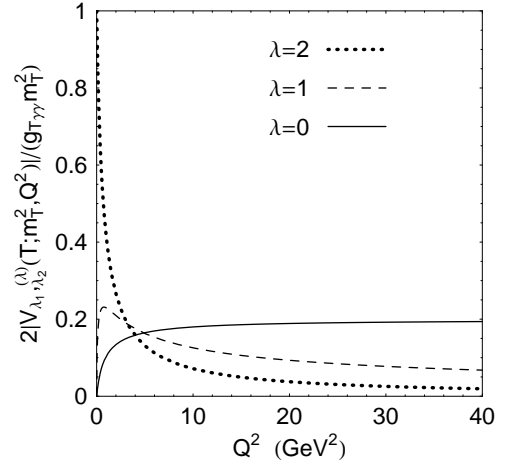}
\caption{\label{V012} The $Q^2$ dependencies of the normalized
vertex functions $2|V^{(\lambda)}_{\lambda_1,
\lambda_2}(T;\,m^2_T,Q^2)| /(g_{T\gamma\gamma}m^2_T)$ calculated
according to Eqs. (\ref{FF2})--(\ref{Norm2}) and (\ref{FFQ2}). For
the $f_2(1270)$ and $a_2 (1320)$ resonances these dependencies
practically coincide because $m_{f_2}\approx m_{a_2}$.}
\end{center}\end{figure}
As is seen from Fig. 2, the main at $Q^2$\,=\,0 vertex function with
helicity $\lambda$\,=\,2 decreases very rapidly with increasing
$Q^2$. For $Q^2$\,$\gtrsim$\,10 GeV$^2$ the vertex function with
helicity $\lambda$\,=\,0 becomes main and close to its asymptotic
value.

The angular distributions in the reactions $\gamma^*\gamma\to
f_2(1270)\to\pi\pi$ \cite{AK86},\, $\gamma^*\gamma\to a_2
(1320)\to\pi^0\eta$,\, and\, $\gamma^*\gamma\to a_2(1320)\to\rho^\pm
\pi^\mp\to\pi^+\pi^-\pi^0$ reshape as $Q^2$ increases with the same
rate.

The differential cross sections for $\gamma^*\gamma\to f_2(1270)\to
\pi\pi$ and $\gamma^*\gamma\to a_2(1320)\to\pi^0\eta$ (integrated
over the azimuth angle of one of the outgoing mesons in the
$\gamma^*\gamma$ center-of-mass system) have the following form:
$\sin^4\theta$ for the tensor meson decays from the helicity
$\lambda$\,=\,$\pm2$ states, $4\cos^2\theta\sin^2\theta$ for
$\lambda$\,=\,$\pm1$, and $\frac{2}{3}(3\cos^2\theta-1)^2$ for
$\lambda$\,=\,0, where $\theta$ is the polar angle of one of the
outgoing mesons. These angular distributions are equally normalized.
Thus, the $\sin^4\theta$ distribution dominating at $Q^2$\,=\,0
should be replaced by the $\frac{2}{3}(3\cos^2\theta-1)^2$
distribution with increasing $Q^2$.


The amplitude of the reaction $\gamma^*\gamma\to a_2(1320)\to
\rho^\pm\pi^\mp\to\pi^+\pi^-\pi^0$ is described by two diagrams and
therefore the corresponding angular distributions for $\lambda
$\,=\,$\pm2$, $\pm1$, and 0 have a rather cumbersome form.
Nevertheless these distributions are sensitive to the $a_2(1320)$
helicity $\lambda$. They are exhaustively represented in Refs.
\cite{BDG84,BBB90,ARG97,AAA97, AAA06}. Here we consider as an
example the contribution of one diagram $\gamma^*\gamma\to
a_2(1320)\to\rho^+\pi^-\to\pi^+\pi^-\pi^0$ only. Then the angular
distributions (integrated over the azimuth angle of the outgoing
$\pi^-$ meson) corresponding to the $\lambda $\,=\,$\pm2$, $\pm1$,
and 0 helicity contributions are
\begin{equation}\label{AD2}
\sin^2\theta_{\rho^+}\sin^2\theta_{\pi^+}(\cos^2\theta_{\rho^+}\sin^2\varphi_{\pi^+}
+\cos^2\varphi_{\pi^+}),
\end{equation}
\begin{equation}\label{AD1}
\sin^2\theta_{\pi^+}[\sin^2\varphi_{\pi^+}(2\cos^2\theta_{\rho^+}-1)^2+\cos^2\varphi_{
\pi^+}\cos^2\theta_{\rho^+}],
\end{equation}
\begin{equation}\label{AD0}
6\sin^2\theta_{\rho^+}\sin^2\theta_{\pi^+}\cos^2\theta_{\rho^+}\sin^2\varphi_{\pi^+},
\end{equation}
respectively, where $\theta_{\rho^+}$ is the polar angle of the
$\rho^+$ in the $\gamma^*\gamma$ center-of-mass system, with the
z-axis along the incident photon direction; the angles
$\theta_{\pi^+ }$ and $\varphi_{\pi^+}$ describe the decay of the
$\rho^+$ in its helicity system; $\varphi_{\pi^+}$ is measured from
the plane defined by the momenta of the $\rho^+$ and photons. As
$Q^2$ increases, the distribution from Eq. (\ref{AD2}) should be
replaced by that from Eq. (\ref{AD0}).

Note that the form factors of a more general form than that in Eq.
(\ref{FFQ2}) may be required for the treatment of real data, for
example,
\begin{equation}\label{gFFQ2} F_T(Q^2)=\frac{1+\xi
Q^2}{(1+Q^2/m^2_\rho)(1+Q^2/m^2_{\rho'})(1+Q^2 /m^2_{\rho''})}
\end{equation} with varying masses $m_{\rho'}$ and
$m_{\rho''}$ and an additional free parameter $\xi$.

Deviations from the above picture are possible in principle since
the tensor meson production in $\gamma^*\gamma$ collisions can be
described in the general case by three independent invariant
amplitudes. However, our scenario is based on the well-established
dominance of the $\lambda$\,=\,$\pm2$ helicity states in the tensor
meson production by two real photons. This allows us to hope that
possible deviations will be small.

Thus, the experiments on the reactions $\gamma^*\gamma\to f_2(1270)$
and $\gamma^*\gamma\to a_2(1320)$ will allow to check the
theoretical predictions about the changing of the dominant helicity
amplitude with increasing $Q^2$. The dynamics of this change can be
tracked by analyzing the angular distributions of the final mesons
in the reactions $\gamma^*\gamma\to f_2(1270)\to\pi\pi$,\,
$\gamma^*\gamma\to a_2(1320)\to\pi^0\eta$,\, and\, $\gamma^*(Q^2)
\gamma\to a_2(1320)\to\rho^\pm\pi^\mp\to\pi^+\pi^-\pi^0$. The
information obtained on three transition form factors, corresponding
to the $\lambda$\,=\,2, 1, and 0 $\gamma^*\gamma$ helicity
amplitudes, would be crucial for the selection of dynamical models
of the $f_2(1270)$ and $a_2(1320)$ resonance production.

We have shown that the transition to the asymptotic behavior of QCD
in the amplitudes $\gamma^*(Q^2)\gamma \to f_2(1270),a_2(1320)$ is
provided by the compensation of the contributions of ground vector
states $\rho$ and $\omega$ in $Q^2$-channel with those of their
radial excitations.

More recently, the Belle Collaboration represented the first data on
the processes $\gamma^*(Q^2)\gamma\to f_2(1270)$ extracted from the
measured differential cross section of the reaction $\gamma^*\gamma
\to f_2(1270)\to\pi^0\pi^0$ for $Q^2$ up to 30 GeV$^2$ \cite{MUW15}.
In Fig. 3, the curves transferred from Fig. 2 are compared with the
Belle data which we multiplied by a factor $(1+Q^2/m^2_{f_2})$ in
order to match the definition of the transition form factors with
$\lambda $\,=\,2, 1, and 0 used in Ref. \cite{MUW15} with our
definition of the normalized vertex functions. The theoretical
curves are in satisfactory agreement with the data.

\begin{figure}  \begin{center}
\includegraphics[width=6.75cm]{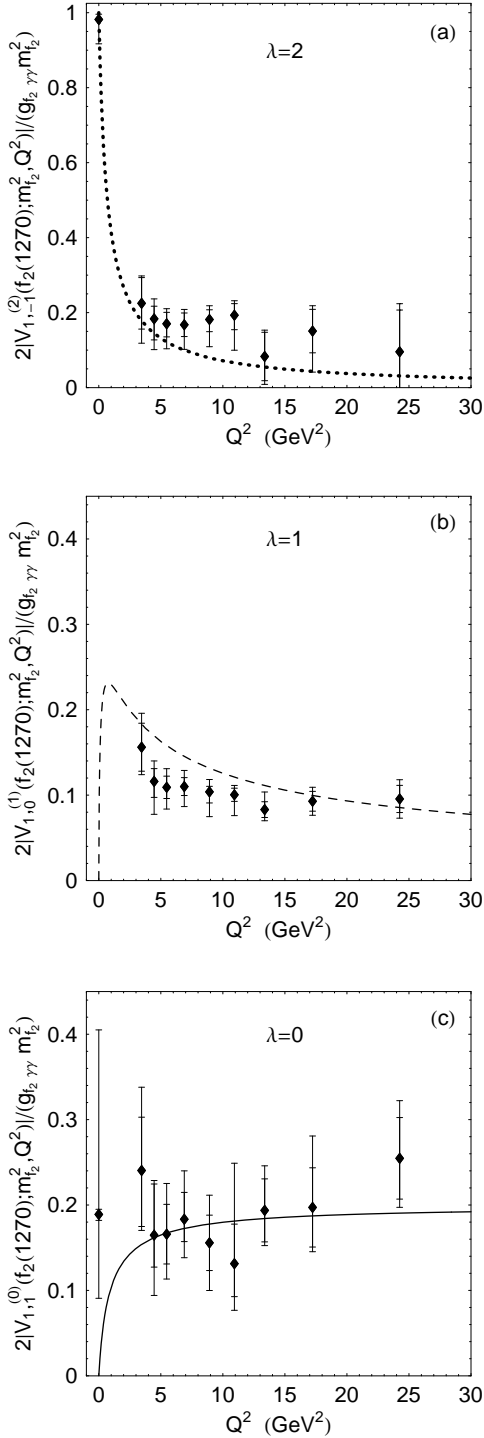}
\caption{\label{V012} Comparison of the $Q^2$ dependencies of the
normalized vertex functions $2|V^{(\lambda)}_{\lambda_1,
\lambda_2}(f_2(1270);\,m^2_{f_2},Q^2)|/(g_{f_2\gamma
\gamma}m^2_{f_2})$, calculated according to Eqs.
(\ref{FF2})--(\ref{Norm2}) and (\ref{FFQ2}), with the Belle data.
\cite{MUW15}. The curves are the same as in Fig. 2. The Belle data
are reduced to our normalization.}\end{center}\end{figure}

\vspace*{0.2cm}

The present work is partially supported by the RFBR grant No.
13-02-00039 and by the Interdisciplinary project No. 102 of the
Siberian Branch of RAS.

\vspace*{0.4cm}


\end{document}